# Evolution of magnetic and microstructural properties of thick sputtered NdFeB films with processing temperature

A. Walther[1,2], K. Khlopkov[3], O. Gutfleisch[3], D. Givord[1], N.M. Dempsey[1]*

[1] Institut Néel, CNRS-UJF, 25 rue de Martyrs, 38042, Grenoble, France

[2] CEA Léti - MINATEC, 17 rue des Martyrs, 38054 Grenoble, France

[3] IFW Dresden, Institute of Metallic Materials, Helmholtzstr. 20, 01069 Dresden, Germany

## Abstract

Ta (100 nm) / NdFeB (5 μm) / Ta (100 nm) films have been deposited onto Si substrates using triode sputtering (deposition rate ~ 18 μm/h). A 2-step procedure was used : deposition at temperatures up to 400°C followed by ex-situ annealing at higher temperatures. Post-deposition annealing temperatures above 650°C are needed to develop high values of coercivity. The duration of the annealing time is more critical in anisotropic samples deposited onto heated substrates than in isotropic samples deposited at lower temperatures. For a given set of annealing conditions (750°C/ 10'), high heating rates (≥ 2000°C / h ) favour high coercivity in both isotropic and anisotropic films. The shape and size of $Nd_2Fe_{14}B$ grains depend strongly on the heating rate.

* Corresponding author. Tel.: 33-4-76887435; fax: 33-4-76881191.
nora.dempsey@grenoble.cnrs.fr

**Introduction**

NdFeB thin films with excellent hard magnetic properties have been deposited by RF sputtering [1], magnetron sputtering [2] and Pulsed Laser Deposition [3]. Thick films of such high performance hard magnetic materials have potential applications in magnetic MEMS (Micro Electro Mechanical Systems) [4]. High quality NdFeB thick films (≤ 300 μm) have been prepared by high rate triode sputtering [5]. These authors were the first to report that out-of-plane texture can be developed in NdFeB films by a 2-step procedure: deposition at a temperature below the crystallization temperature followed by a high temperature post-deposition anneal. While samples deposited at room temperature were magnetically isotropic, samples deposited at higher temperatures were anisotropic and the degree of texture was found to increase with substrate temperature [5]. We have studied the evolution in both magnetic and microstructural properties in 5 μm thick films, produced using such a 2-step procedure, as a function of substrate temperature [6]. In this study, the influence of post-deposition annealing conditions on the magnetic and microstructural properties of NdFeB films deposited under different conditions is assessed.

**Experiment**

A $Nd_{16.8}Fe_{74.7}B_{8.5}$ target was triode sputtered in an Ar plasma at a power of 1000 W. 5 μm thick NdFeB layers, with 100 nm thick Ta buffer and capping layers, were deposited at a rate of 18μm/h onto Si. The substrate was either not furnace heated (we label this "cold" deposition, although the substrate temperature rose to 230 °C during deposition) or maintained at a fixed temperature of 400 °C. Samples were then annealed ex-situ in a vacuum of $10^{-5}$ mbar. The heating rate, annealing temperature and annealing time were varied.

Magnetic properties were measured at 300 K with an Oxford VSM. Microstructures were investigated with a high resolution LEO FEG SEM (GEMINI 1530).

**Results and discussion**

Annealing treatments at different temperatures have been carried out on "cold" deposited NdFeB films. These samples were rapidly heated (from room temperature to the annealing temperature in approximately 40 seconds) and maintained at the annealing temperature for 10 minutes. Such films do not show any preferential texture and their in-plane hysteresis loops are compared in figure 1. The highest coercivity ($\mu_0H_c$ = 1.95 T) was obtained for the maximum annealing temperature (850 °C). Note that coercivity is significantly increased between the annealing temperatures of 650 °C and 675 °C. Considering that the x-ray diffraction peaks of $Nd_2Fe_{14}B$ are clearly resolved for all samples (data not shown), we tentatively attribute this large increase in coercivity to the redistribution of a Nd-rich intergranular phase leading to a more homogeneous and complete covering, and thus better decoupling, of the $Nd_2Fe_{14}B$ grains.

The influence of annealing time was assessed on both the "cold" and hot deposited films. Samples were rapidly heated to 750 °C and annealed for times of 0 to 60 minutes. "Cold" deposited films show high coercivity even for the shortest annealing time and the coercivity varies little over the time scale studied, though a slight decrease is observed for the longest annealing time of 60 minutes (figure 2a). On the contrary, the coercivity of hot deposited films, which show high crystallographic texture (c-axis perpendicular to the film plane), increases significantly with annealing time (figure 2b). In this case, a maximum coercivity of 1.7 T was achieved for the longest annealing time.

The influence of heating rate was also assessed. Both hot and "cold" deposited films were annealed at 750 °C for 10 minutes with heating rates of 500 °C/h, 1000 °C/h, 2000 °C/h and a rapid heating (RT to annealing temperature in 40 seconds, i.e. 65000 °C/h) as used above. For both film types, magnetic properties are strongly dependent on the heating rate and a fast heating rate promotes the development of coercivity. However, differences can be noticed between the behaviour of "cold" deposited (figure 3(a)) and hot deposited films (figure 3(b)). Whereas, for "cold" deposited films, a heating rate of 1000°C/h yields the same magnetic properties as for 500°C/h, coercivity is higher for the faster heating rate in hot deposited films. Moreover, contrary to "cold" deposited films, annealing at 2000°C/h yields a slightly higher coercivity than a rapid heating of hot deposited ones.

The shape and size of the $Nd_2Fe_{14}B$ grains are found to depend strongly on the heating rate for both film types. While rapidly heated films have equiaxed grains throughout the film thickness, films heated at slower rates have a mixed structure, with columnar grains in the lower film section and equiaxed grains in the upper film section (figure 4).

Plane view images of the top of the films show that the equiaxed grains in the films heated at slow rates are much larger than those formed in the rapidly heated films (figure 5). In our previous study, it was shown that both the size and shape of $Nd_2Fe_{14}B$ grains depend strongly on the substrate temperature during deposition, and a surprising result was that samples with very different grain sizes (in the range 0.1 – 1 μm) and shapes (equiaxed vs columnar) had very similar values of coercivity [6]. Thus, the difference in the size and shape of grains in samples heated at different rates is not sufficient to explain the differences in coercivities observed. These differences may be due to differences in the distribution of a Nd-rich intergranular phase and secondary phases. TEM studies are now underway.

**Conclusion**

5μm thick NdFeB films have been deposited at a deposition rate of 18 μm/h and excellent hard magnetic properties were achieved ($\mu_0H_c \leq 1.95T$) following post-deposition annealing. Films deposited onto heated substrates are highly textured. The threshold temperature for developing coercivity is higher than the crystallization temperature of the $Nd_2Fe_{14}B$ phase and is in the range of the eutectic melting temperature of the Nd-rich intergranular phase. For the post-deposition annealing conditions studied, high heating rates favour high coercivity.

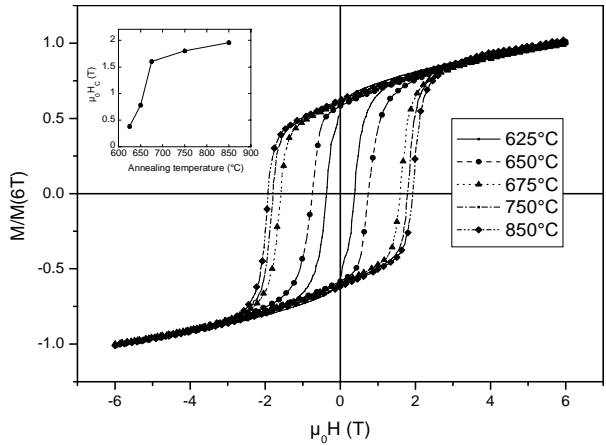

Figure 1 : In-plane VSM measurements of Si/Ta/NdFeB(5µm)/Ta films "cold" deposited and subsequently annealed at different temperatures for 10 minutes (inset : coercivity vs annealing temperature).

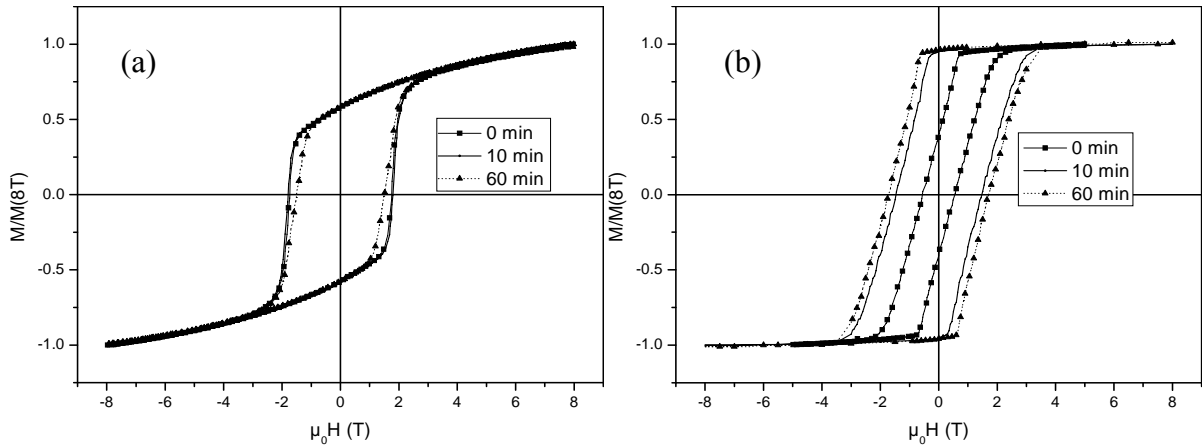

Figure 2 : VSM measurements of Si/Ta/NdFeB(5µm)/Ta films deposited "cold" ((a), in-plane loops) and at 400°C ((b) out-of-plane loops non corrected for demagnetisation), and subsequently annealed at 750 °C for different times.

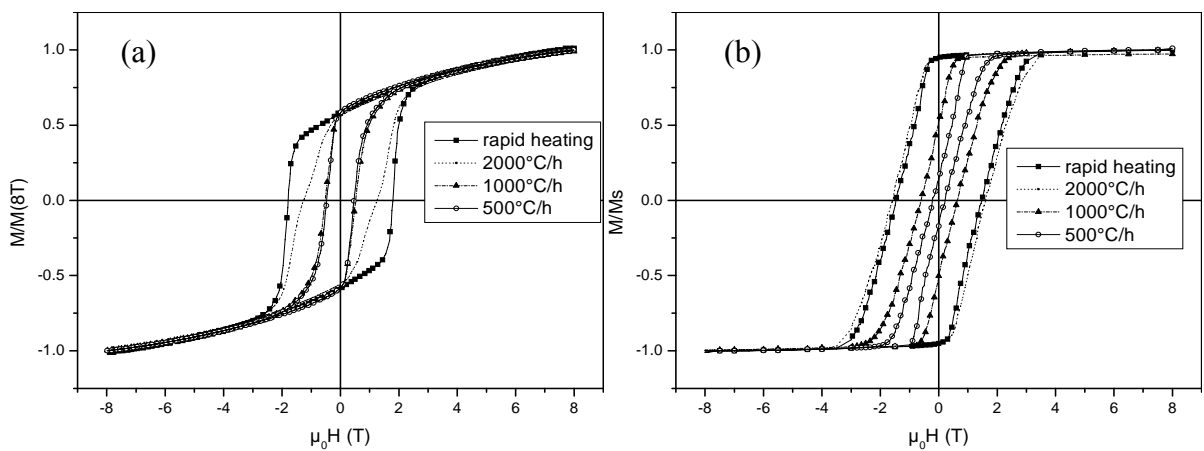

Figure 3 : VSM measurements of Si/Ta/NdFeB(5µm)/Ta layers deposited "cold" ((a), in-plane loops) or at 400°C ((b), out-of-plane loops, non corrected for demagnetisation) and annealed at 750 °C for 10 min with different heating rates.

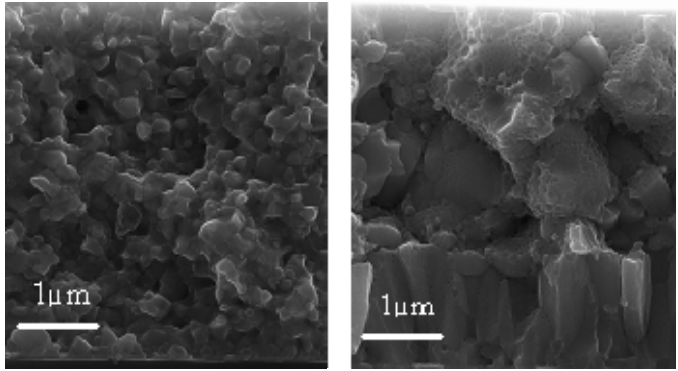

Figure 4 : SEM cross sectional view (in-lens detector) of the fracture surface of "cold" deposited NdFeB films annealed at 750 °C for 10' with a rapid heating rate (left) and a heating rate of 1000 °C/h (right).

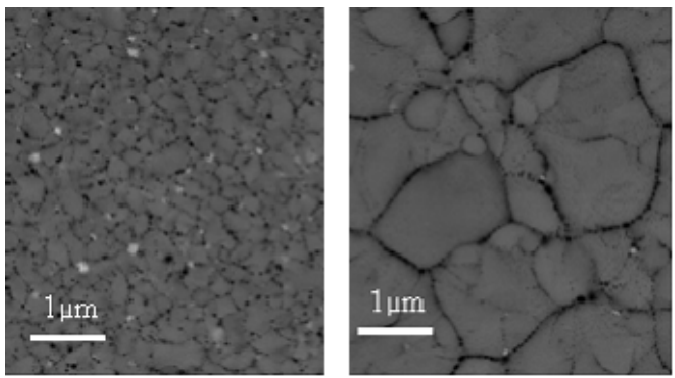

Figure 5 : SEM plane view (QBSD detector) of polished surfaces of "cold" deposited NdFeB films annealed at 750 °C for 10' with a rapid heating rate (left) and a heating rate of 1000 °C/h (right). The white contrast is attributed to a Nd-rich phase.